\documentclass[12pt,prd,tightenlines,nofootinbib]{revtex4}
\usepackage{bm}
\usepackage{graphics}
\usepackage{rotating}
\usepackage{epsfig}
\begin{document}
\title{
Relativistic description of the $\Xi_b$ baryon semileptonic decays}  
\author{R. N. Faustov}
\author{V. O. Galkin}
\affiliation{Institute of Cybernetics and  Informatics in Education, FRC CSC RAS,
  Vavilov Street 40, 119333 Moscow, Russia}

\begin{abstract}
Semileptonic decays of the $\Xi_b$ baryon are studied in the framework of the
relativistic quark-diquark model based on the quasipotential approach. The weak decay form factors are calculated
with the comprehensive account of all relativistic effects without
employing nonrelativistic and heavy quark expansions. On this
basis differential and total decay rates as well as different
asymmetry parameters are calculated for the heavy-to-heavy $\Xi_b\to \Xi_c\ell\nu_\ell$
and heavy-to-light $\Xi_b\to \Lambda\ell\nu_\ell$ semileptonic
decays. Predictions for the ratios of such decays involving $\tau$
lepton and muon are presented.   
\end{abstract}
\vspace*{0.2cm}

\maketitle

\section{Introduction}

The investigation of the semileptonic decays of bottom baryons
represents a very interesting and important problem. Indeed, their
study provides an independent determination of the
Cabibbo-Kobayashi-Maskawa (CKM) matrix elements $|V_{cb}|$ and
$|V_{ub}|$ and thus can help to better understand the origin of the disagreement of
their values determined from inclusive and exclusive $B$ meson
decays \cite{pdg}. Such decays can be also
used to verify the lepton flavour universality,  indications of
which violation were reported in the semileptonic $B$ meson decays governed by
the $b\to c$ quark transitions (for recent review see \cite{bifani}
and references therein). The discrepancy between predictions of the
Standard Model (SM) and experimental data was observed for the ratios of
the branching fractions $R_D$ and $R_{D^*}$ of the semileptonic $B$ meson decays to $D$
and $D^*$ mesons, respectively, involving
$\tau$ lepton and muon. The combined excess of the
measured ratios over the SM prediction is about $3.6\sigma$ \cite{bifani}. Recently the
LHCb Collaboration \cite{lhcbjp} measured the ratio $R_{J/\Psi}$ of semileptonic
$B_c$ meson decays to $J/\Psi$ with $\tau$ lepton and muon which
exceeds the SM prediction by more than $2\sigma$ \cite{bifani,tiks}.

In our paper \cite{Lambdabsl} we comprehensively investigated
semileptonic decays of $\Lambda_b$ baryons in the framework of the
relativistic quark-diquark model based on the quasipotential approach. The
explicit expressions for the decay form factors were obtained in terms
of the
overlap integrals over the baryon wave functions. Baryons were treated
as quark-diquark composite systems. All relativistic effects including
contributions of the intermediate negative-energy states and wave
function transformations from the rest to the moving reference frame were
systematically taken into account. To achieve this goal we do not use
either nonrelativistic or heavy quark expansions. As a
result the obtained formulas are valid in the whole range of the momentum
transfer $q^2$ both for the heavy-to-heavy
($b\to c$ weak transitions)
and heavy-to-light ($b \to u$ weak transitions) semileptonic decays of bottom
baryons. The calculated form factors of the $\Lambda_b$ baryon
transitions were used for obtaining predictions for the differential
and total decay rates and different asymmetry parameters.  Good
agreement of theoretical results with experimental data was found.

In the present paper we extend the previous analysis to the semileptonic
decays of the $\Xi_b$ baryon. Such decay are significantly less studied
both theoretically and experimentally. However there are good chances
that they will be soon observed at LHC.  

\section{Weak decay form factors}

The hadronic matrix elements of the vector and axial vector weak
currents for the semileptonic decay $\Xi_b\to \Xi_c(\Lambda)$  are parametrized  in terms of
six invariant form factors \cite{giklsh}
\begin{eqnarray}
  \label{eq:ff}
  \langle \Xi_c(\Lambda)(p',s')|V^\mu|\Xi_b(p,s)\rangle&=& \bar
  u_{\Xi_c(\Lambda)}(p',s')\Bigl[f_1^V(q^2)\gamma^\mu-f_2^V(q^2)i\sigma^{\mu\nu}\frac{q_\nu}{M_{\Xi_b}}\cr&&+f_3^V(q^2)\frac{q^\mu}{M_{\Xi_b}}\Bigl]
u_{\Xi_b}(p,s),\cr
 \langle \Xi_c(\Lambda)(p',s')|A^\mu|\Xi_b(p,s)\rangle&=& \bar
  u_{\Xi_c(\Lambda)}(p',s')[f_1^A(q^2)\gamma^\mu-f_2^A(q^2)i\sigma^{\mu\nu}\frac{q_\nu}{M_{\Xi_b}}\cr&&+f_3^A(q^2)\frac{q^\mu}{M_{\Xi_b}}\Bigl]
\gamma_5 u_{\Xi_b}(p,s),\qquad 
\end{eqnarray}
where $M_{B}$ and  $u_B(p,s)$ are masses and Dirac spinors of
the $B$  baryons ($\Xi_b,\Xi_c,\Lambda$), $q=p'-p$.

The expressions for these form factors as the overlap integrals of
baryon wave functions with the systematic account of the relativistic
effects are given in Ref.~\cite{Lambdabsl}. They were obtained without
employing the heavy quark expansion both for initial and final
baryons. Therefore we can apply them for the calculation of the
heavy-to-heavy $\Xi_b\to\Xi_c\ell\nu_\ell$ and heavy-to-light
$\Xi_b\to\Lambda\ell\nu_\ell$ decay form factors. For the numerical calculations
we use the baryon wave functions obtained while studying their
spectroscopy \cite{hbarregge,sbar}.  

\begin{figure}
\centering
  \includegraphics[width=8cm]{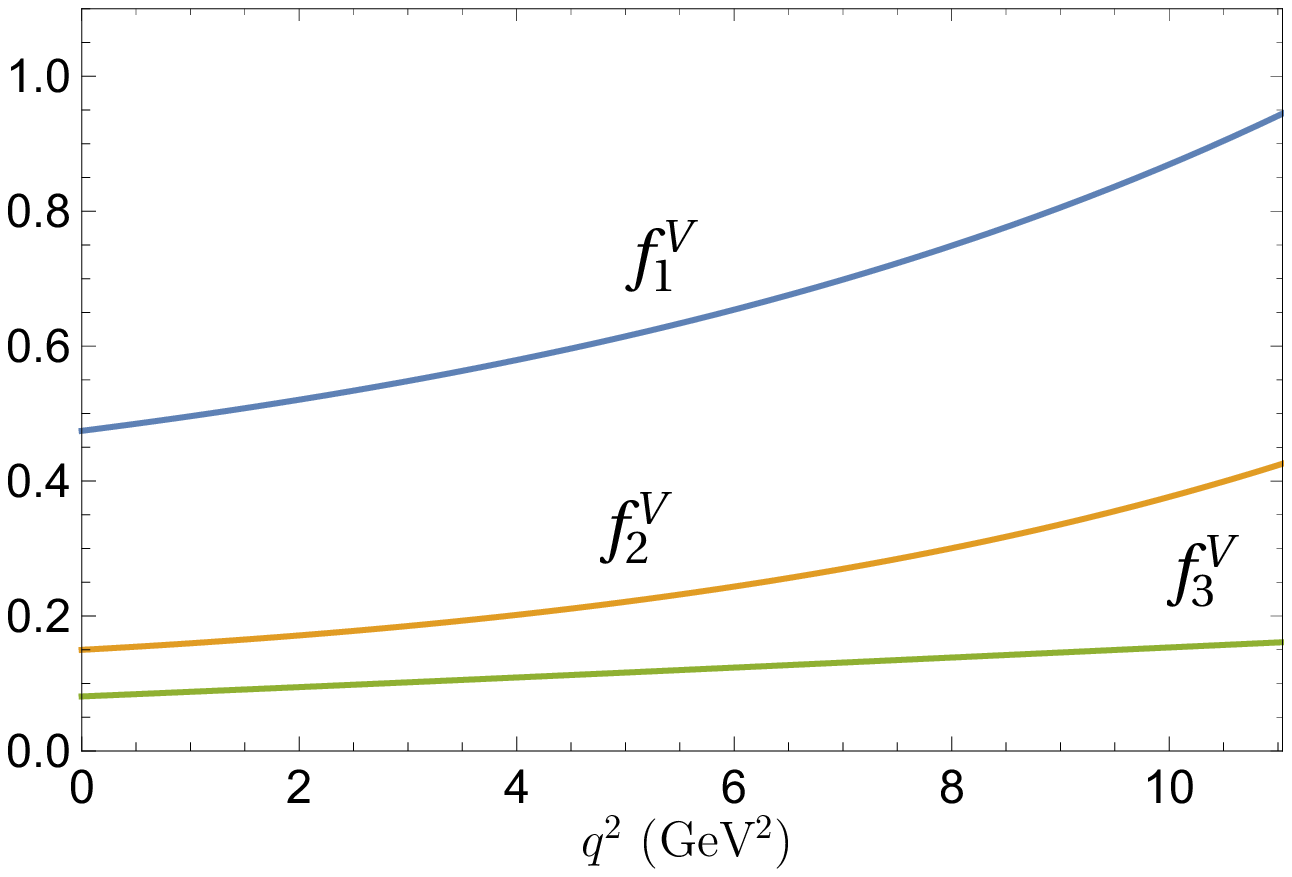}\ \
 \ \includegraphics[width=8cm]{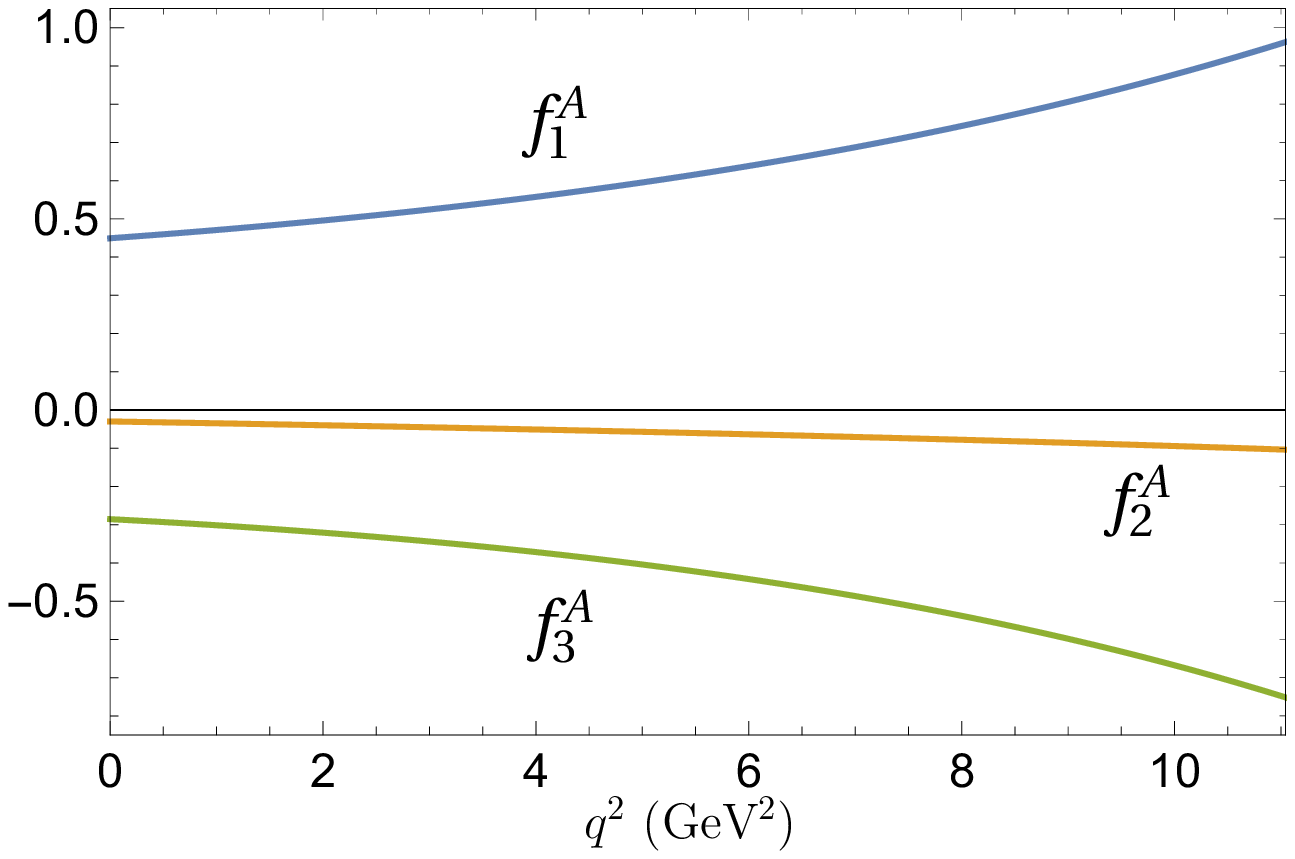}\\
\caption{Form factors of the weak $\Xi_b\to \Xi_c$ transitions.    } 
\label{fig:ffXibXic}
\end{figure}

\begin{figure}
\centering
  \includegraphics[width=8cm]{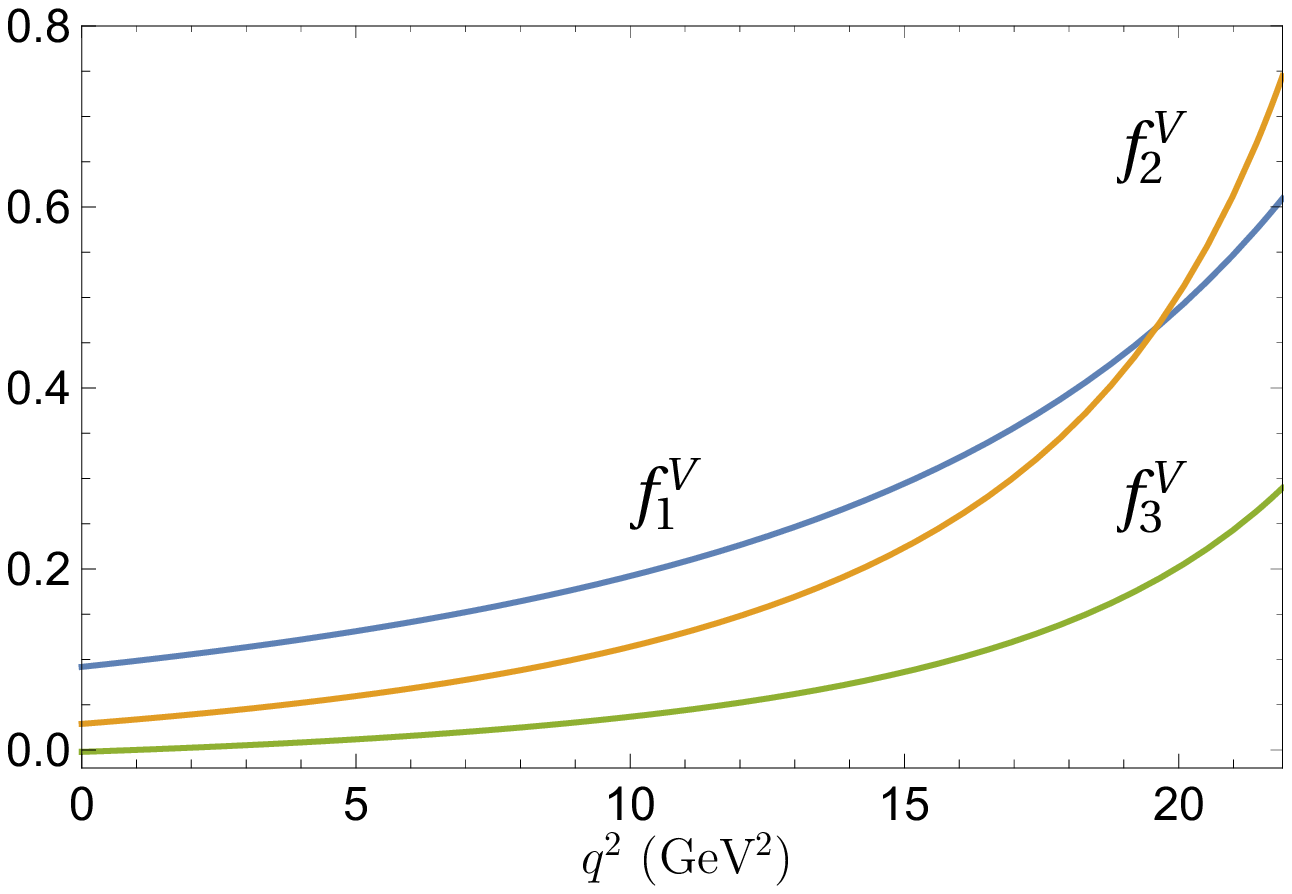}\ \
 \ \includegraphics[width=8cm]{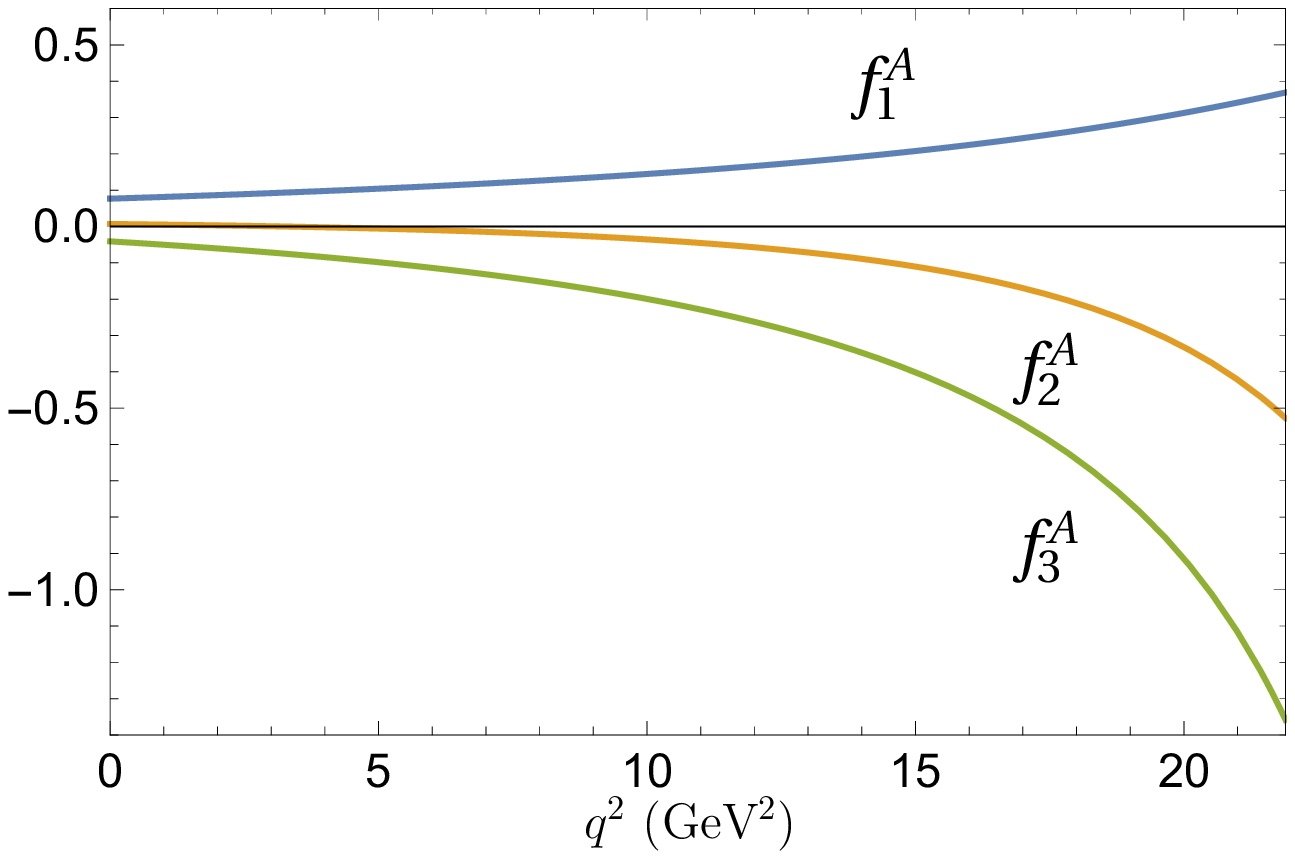}\\
\caption{Form factors of the weak $\Xi_b\to \Lambda$ transitions.    } 
\label{fig:ffXibLambda}
\end{figure}

We found that the numerically calculated form factors can be
approximated with high accuracy by the following analytic expression
\begin{equation}
  \label{fitff}
  f(q^2)= \frac{1}{{1-q^2/{M_{\rm pole}^2}}} \left\{ a_0 + a_1 z(q^2) +
    a_2 [z(q^2)]^2 \right\},
\end{equation}
where the variable 
\begin{equation}
z(q^2) = \frac{\sqrt{t_+-q^2}-\sqrt{t_+-t_0}}{\sqrt{t_+-q^2}+\sqrt{t_+-t_0}},
\end{equation}
here $t_+=(M_B+M_\pi)^2$ and $t_0 = q^2_{\rm max} = (M_{\Xi_b} - M_{\Xi_c(\Lambda)})^2$.  The pole
masses have the following  values:\\
a) for $\Xi_b\to\Xi_c$ transitions\\
$M_{\rm
  pole}\equiv M_{B_c^*}=6.333$ GeV for $f_{1,2}^V$; $M_{\rm
  pole}\equiv M_{B_{c1}}=6.743$ GeV for $f_{1,2}^A$;\\ $M_{\rm
  pole}\equiv M_{B_{c0}}=6.699$ GeV for $f_{3}^V$;  $M_{\rm
  pole}\equiv M_{B_c}=6.275$ GeV for $f_{3}^A$;\\
b) for $\Xi_b\to\Lambda$ transitions\\ $M_{\rm
  pole}\equiv M_{B^*}=5.325$ GeV for $f_{1,2}^V$; $M_{\rm
  pole}\equiv M_{B_{1}}=5.723$ GeV for $f_{1,2}^A$;\\ $M_{\rm
  pole}\equiv M_{B_{0}}=5.749$ GeV for $f_{3}^V$;  $M_{\rm
  pole}\equiv M_{B}=5.280$ GeV for $f_{3}^A$.\\
We take the masses of the excited $B_c$ and $B$ mesons from our previous
study of their spectroscopy \cite{bcmass,hmmass}.
The fitted values of the parameters $a_0$, $a_1$, $a_2$ as well as the
values of form factors at maximum $q^2=0$ and zero recoil $q^2=q^2_{\rm
  max}$ are given in Tables~\ref{ffXibXic}, \ref{ffXibLambda}. The difference of the fitted
form factors from the calculated ones does not exceed 0.5\%. Our model form factors
are plotted in Figs.~\ref{fig:ffXibXic}, \ref{fig:ffXibLambda}. We roughly estimate the total
uncertainty of our form factor calculation to be about 5\%.

\begin{table}
\caption{Form factors of the weak $\Xi_b\to \Xi_c$ transitions. }
\label{ffXibXic}
\begin{ruledtabular}
\begin{tabular}{ccccccc}
& $f^V_1(q^2)$ & $f^V_2(q^2)$& $f^V_3(q^2)$& $f^A_1(q^2)$ & $f^A_2(q^2)$ &$f^A_3(q^2)$\\
\hline
$f(0)$          &0.474 &$0.150$ & $0.081$ & 0.449 & $-0.030$&$-0.285$\\
$f(q^2_{\rm max})$&0.945  &$0.426$ & $0.161$ & 0.962& $-0.104$& $-0.752$ \\
$a_0$      &$0.684$&$0.308$& $0.121$& $0.729$ &$-0.078$&  $-0.541$\\
$a_1$      &$-5.16$&$-4.18$&$-0.315$&$-7.11$&$0.775$&$6.93$\\
$a_2$      &$28.0$&$25.9$& $-5.81$& $41.5$ &$0.372$&  $-44.9$\\
\end{tabular}
\end{ruledtabular}
\end{table}

\begin{table}
\caption{Form factors of the weak $\Xi_b\to \Lambda$ transitions. }
\label{ffXibLambda}
\begin{ruledtabular}
\begin{tabular}{ccccccc}
& $f^V_1(q^2)$ & $f^V_2(q^2)$& $f^V_3(q^2)$& $f^A_1(q^2)$ & $f^A_2(q^2)$ &$f^A_3(q^2)$\\
\hline
$f(0)$          &0.092 &$0.029$ & $-0.002$ & 0.077 & $0.007$&$-0.041$\\
$f(q^2_{\rm max})$&0.609  &$0.745$ & $0.290$ & 0.369& $-0.528$& $-1.36$ \\
$a_0$      &$0.139$&$0.170$& $0.098$& $0.122$ &$-0.175$&  $-0.292$\\
$a_1$      &$0.136$&$-0.368$&$-0.323$&$0.016$&$0.865$&$0.554$\\
$a_2$      &$-0.845$&$-0.180$& $0.059$& $-0.470$ &$-0.947$&  $0.630$\\
\end{tabular}
\end{ruledtabular}
\end{table}

\section{Semileptonic $\Xi_b\to\Xi_c\ell\nu_\ell$ and 
$\Xi_b\to \Lambda\ell\nu_\ell$ decays}

Now we can use the obtained form factors for the calculation of the
differential and total semileptonic decay rates, different asymmetry parameters
and other observables. To achieve this goal it is convenient to use
the helicity formalism \cite{giklsh}. The helicity amplitudes are
expressed through the decay form factors by the following relations:
\begin{eqnarray}
  \label{eq:haa}
 H_{+\frac12 0}^{V,A}&=&\frac{\sqrt{(M_{\Xi_b}\mp M_{\Xi_c(\Lambda)})^2-q^2}}{\sqrt{q^2}}
  \bigg[(M_{\Xi_b}\pm M_{\Xi_c(\Lambda)}) f_1^{V,A}(q^2)\pm \frac{q^2}{M_{\Xi_b}} f_2^{V,A}(q^2)\bigg],
\cr
H_{+\frac12 +1}^{V,A}&=&\sqrt{2[(M_{\Xi_b}\mp M_{\Xi_c(\Lambda)})^2-q^2]}
  \bigg[f_1^{V,A}(q^2)\pm \frac{M_{\Xi_b}\pm M_{\Xi_c(\Lambda)}}{M_{\Xi_b}}f_2^{V,A}(q^2)\bigg],
\cr
 H_{+\frac12 t}^{V,A}&=&\frac{\sqrt{(M_{\Xi_b}\pm M_{\Xi_c(\Lambda)})^2-q^2}}{\sqrt{q^2}}
  \bigg[ (M_{\Xi_b}\mp M_{\Xi_c(\Lambda)}) f_1^{V,A}(q^2)\pm \frac{q^2}{M_{\Xi_b}} f_3^{V,A}(q^2)\bigg]. 
\end{eqnarray}
The amplitudes for negative values of the helicities can be obtained
using the relation
$$H^{V,A}_{-\lambda',\,-\lambda_W}=\pm H^{V,A}_{\lambda',\, \lambda_W}.$$
The total helicity amplitude for the
$V-A$ current is given by
\begin{equation}
\label{ha}
H_{\lambda',\, \lambda_W}=H^{V}_{\lambda',\, \lambda_W}
-H^{A}_{\lambda',\, \lambda_W}.
\end{equation}
The helicity structures entering the differential decay rates and angular distributions are expressed in
terms of the total helicity amplitudes (\ref{ha}) by
\begin{eqnarray}
  \label{eq:hhs}
  {\cal H}_U(q^2)&=&|H_{+1/2,+1}|^2+|H_{-1/2,-1}|^2,\cr
{\cal H}_L(q^2)&=&|H_{+1/2,0}|^2+|H_{-1/2,0}|^2,\cr
{\cal H}_S(q^2)&=&|H_{+1/2,t}|^2+|H_{-1/2,t}|^2,\cr
{\cal H}_{SL}(q^2)&=&{\rm Re}(H_{+1/2,0}H_{+1/2,t}^\dag+H_{-1/2,0}H_{-1/2,t}^\dag),\cr
  {\cal H}_P(q^2)&=&|H_{+1/2,+1}|^2-|H_{-1/2,-1}|^2,\cr
{\cal H}_{L_P}(q^2)&=&|H_{+1/2,0}|^2-|H_{-1/2,0}|^2,\cr
{\cal H}_{S_P}(q^2)&=&|H_{+1/2,t}|^2-|H_{-1/2,t}|^2.
\end{eqnarray}

\begin{figure}
  \centering
 \includegraphics[width=8cm]{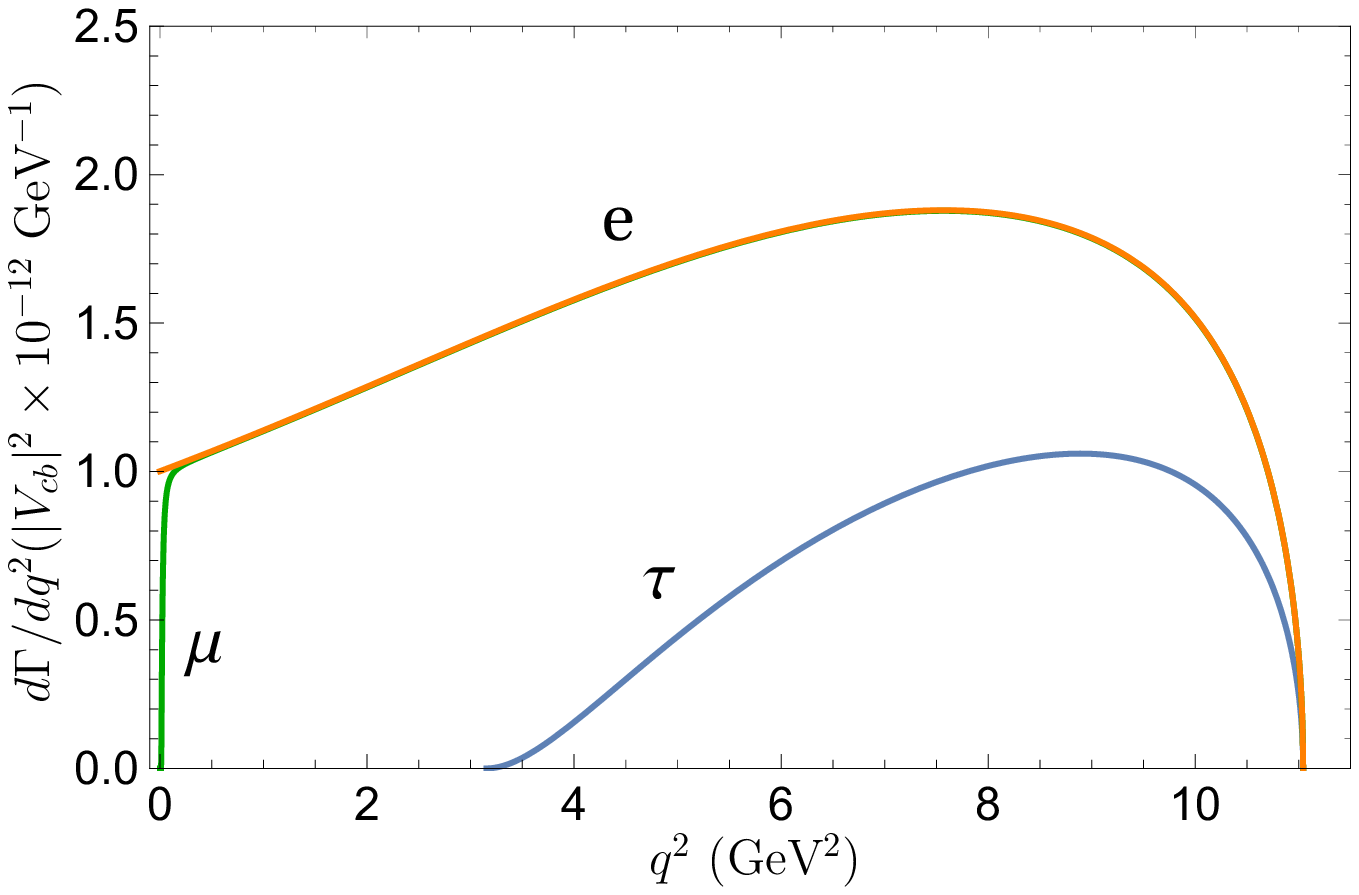}\ \
 \  \includegraphics[width=8cm]{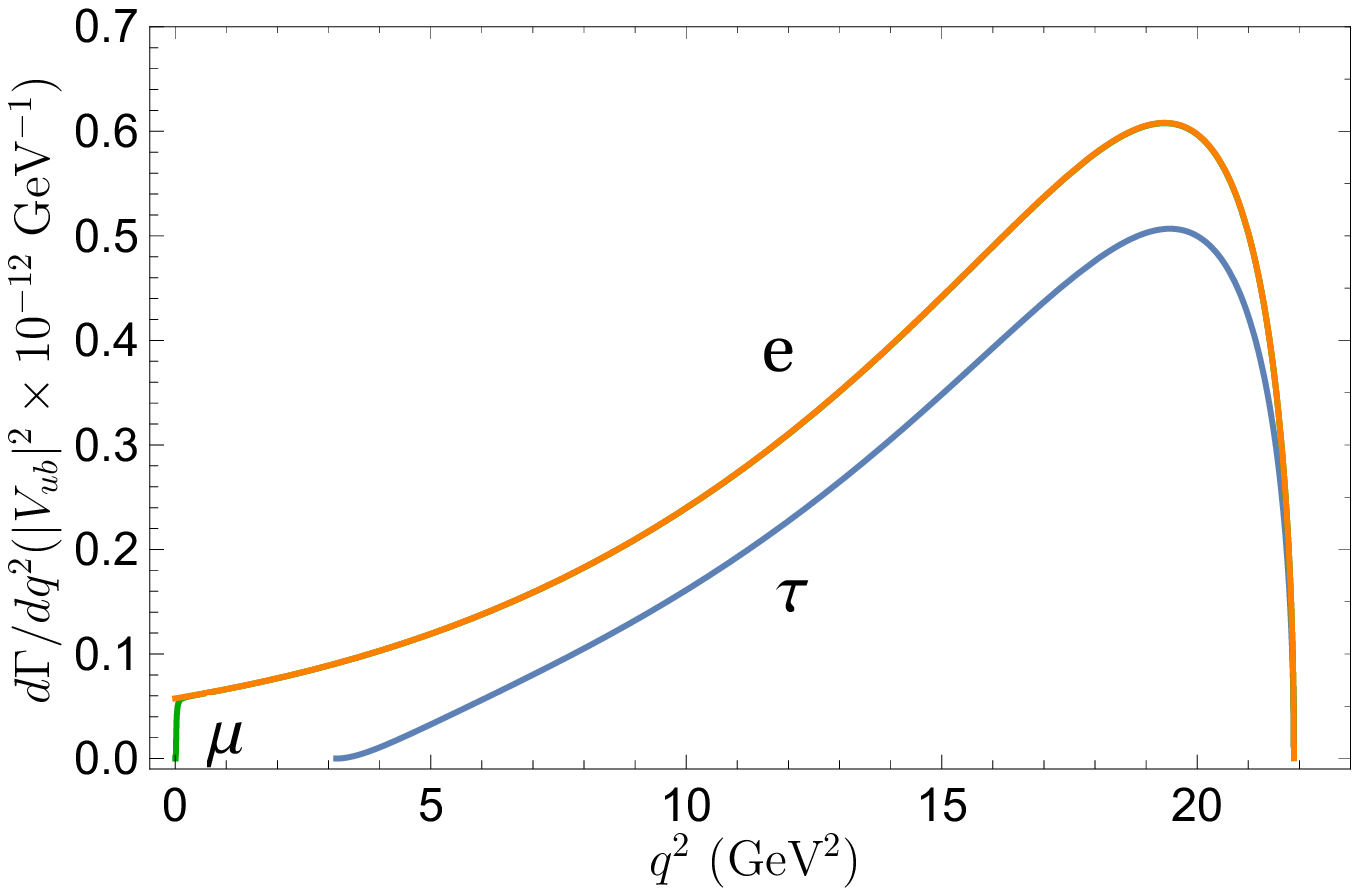}

  \caption{The differential decay rates  of the
    $\Xi_b\to \Xi_c\ell\nu_\ell$ (left) and $\Xi_b\to \Lambda\ell\nu_\ell$ (right)
    semileptonic decays. }
  \label{fig:brXib}
\end{figure}

Then the differential decay rate can be presented by \cite{giklsh}
\begin{equation}
  \label{eq:dgamma}
  \frac{d\Gamma(\Xi_b\to \Xi_c(\Lambda)\ell\bar\nu_\ell)}{dq^2}=\frac{G_F^2}{(2\pi)^3}
  |V_{qb}|^2
  \frac{\lambda^{1/2}(q^2-m_\ell^2)^2}{48M_{\Xi_b}^3q^2}{\cal H}_{tot}(q^2),
\end{equation}
where $G_F$ is the Fermi constant, $V_{qb}$ is the CKM matrix element ($q=c,u$),
 $\lambda\equiv
\lambda(M_{\Xi_b}^2,M_{\Xi_c(\Lambda)}^2,q^2)=M_{\Xi_b}^4+M_{\Xi_c(\Lambda)}^4+q^4-2(M_{\Xi_b}^2M_{\Xi_c(\Lambda)}^2+M_{\Xi_c(\Lambda)}^2q^2+M_{\Xi_b}^2q^2)$,
and $m_\ell$ is the lepton mass ($\ell=e,\mu,\tau$),
\begin{equation}
 \label{eq:hh}
 {\cal H}_{tot}(q^2)=[{\cal H}_U(q^2)+{\cal H}_L(q^2)] \left(1+\frac{m_\ell^2}{2q^2}\right)+\frac{3m_\ell^2}{2q^2}{\cal H}_S(q^2) .
\end{equation}
It is plotted in Fig.~\ref{fig:brXib} for  the $\Xi_b\to \Xi_c\ell\nu_\ell$ (left) and $\Xi_b\to \Lambda\ell\nu_\ell$ (right) semileptonic decays.

\begin{figure}
  \centering
 \includegraphics[width=8cm]{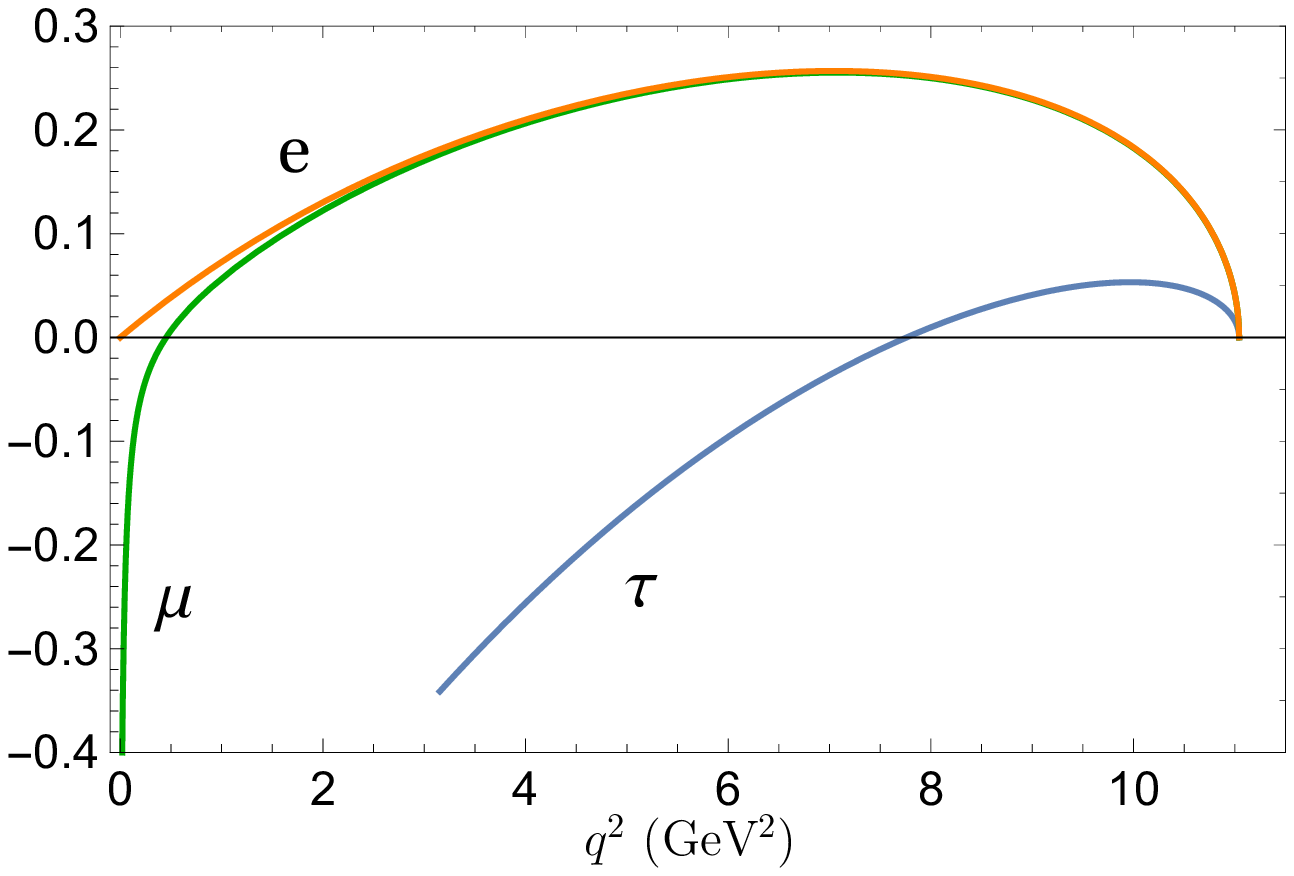}\ \
 \  \includegraphics[width=8cm]{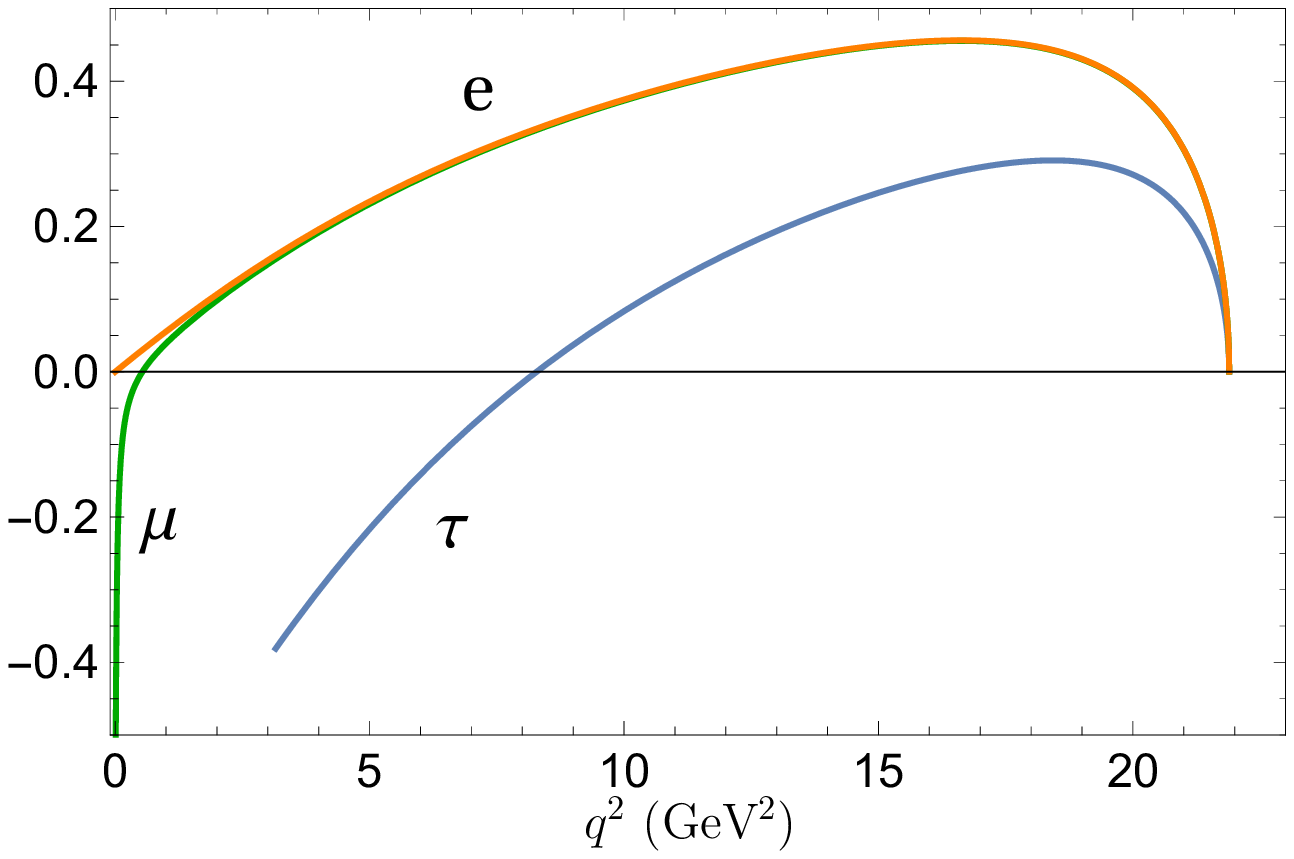}

  \caption{Predictions for the forward-backward asymmetries $A_{FB}(q^2)$ in the
    $\Xi_b\to \Xi_c\ell^-\nu_\ell$ (left) and $\Xi_b\to \Lambda\ell^-\nu_\ell$ (right)
    semileptonic decays. }
  \label{fig:afbXib}
\end{figure}

\begin{figure}
  \centering
 \includegraphics[width=8cm]{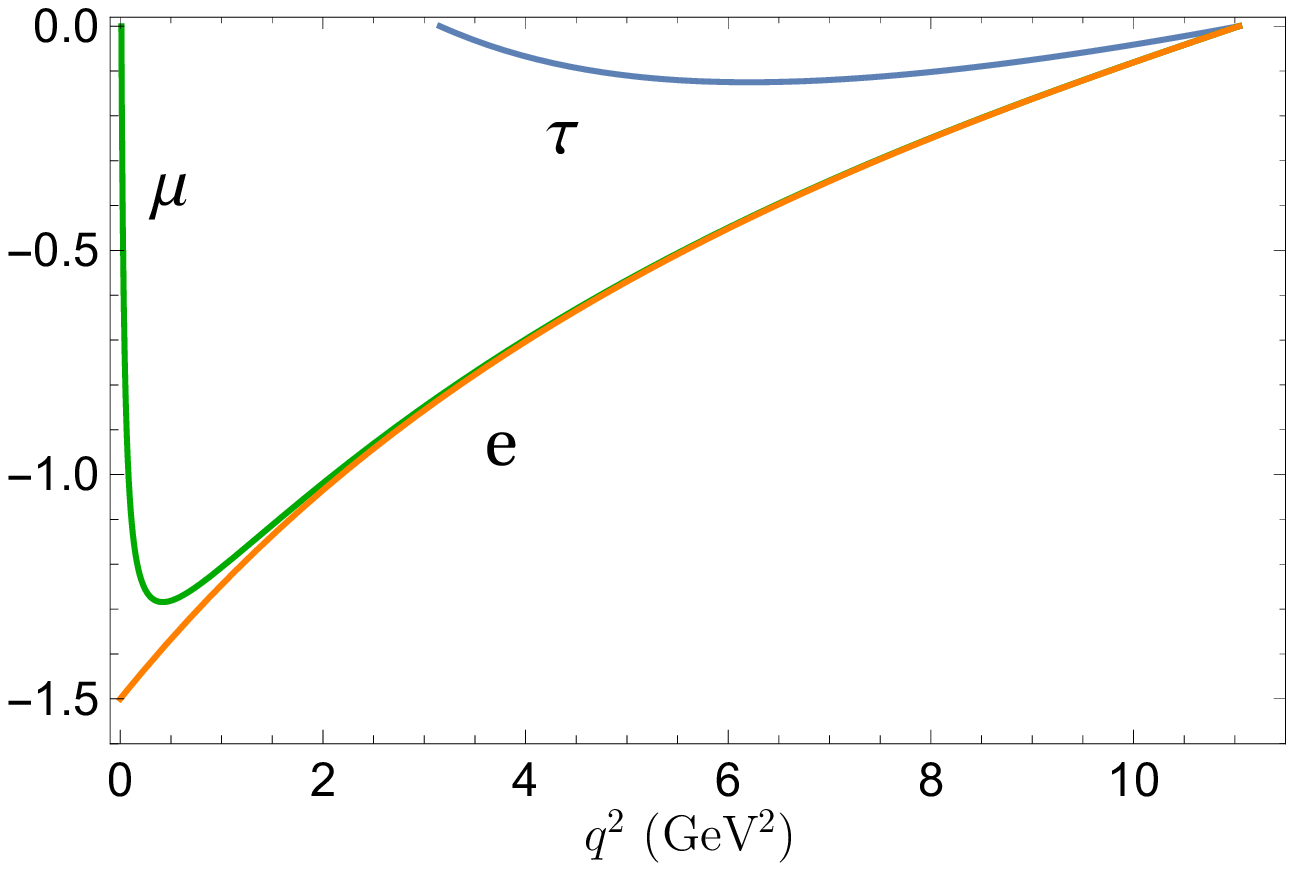}\ \
 \  \includegraphics[width=8cm]{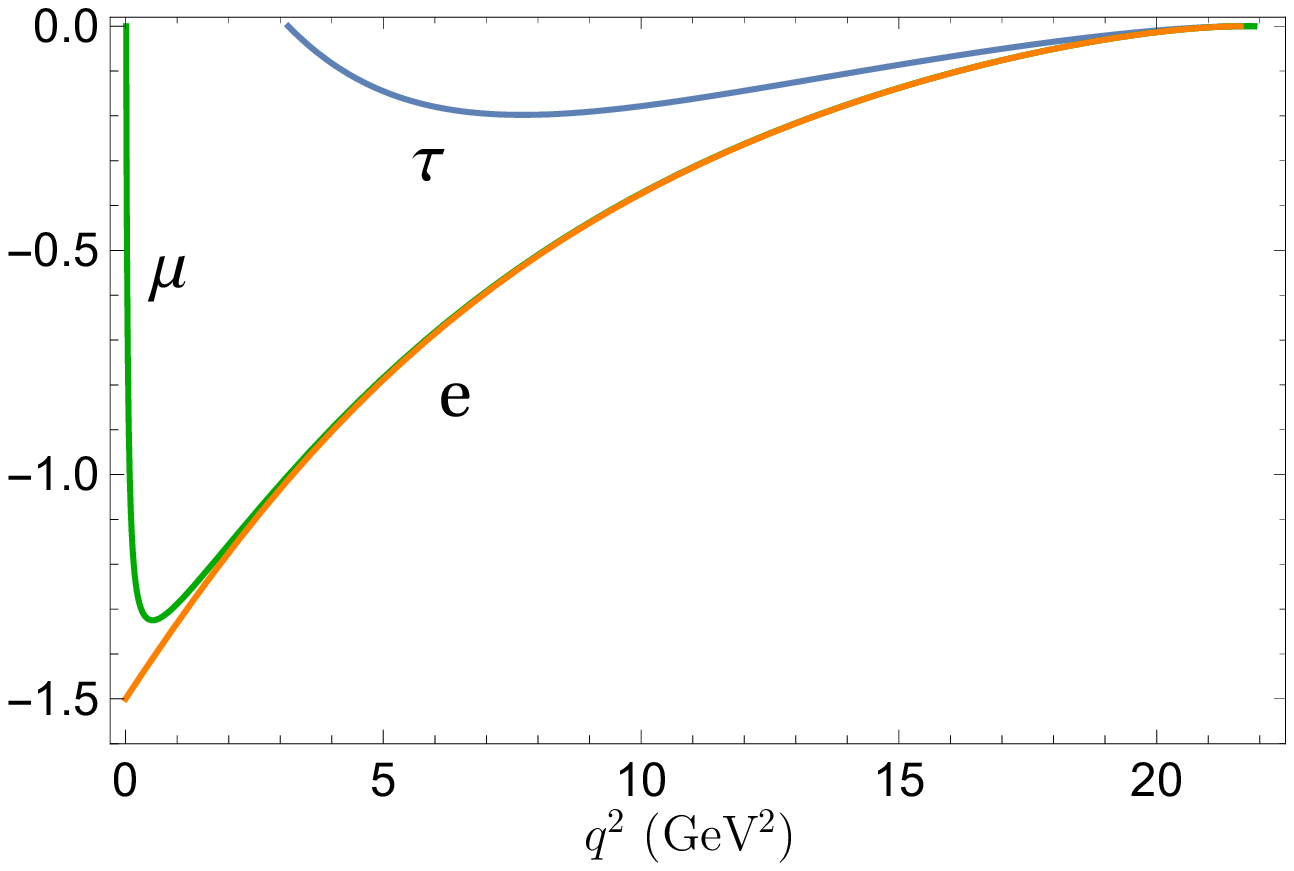}

  \caption{Predictions for the convexity parameter  $C_F(q^2)$ in the
    $\Xi_b\to \Xi_c\ell\nu_\ell$ (left) and $\Xi_b\to \Lambda\ell\nu_\ell$ (right)
    semileptonic decays. }
  \label{fig:cfXib}
\end{figure}

\begin{figure}
  \centering
 \includegraphics[width=8cm]{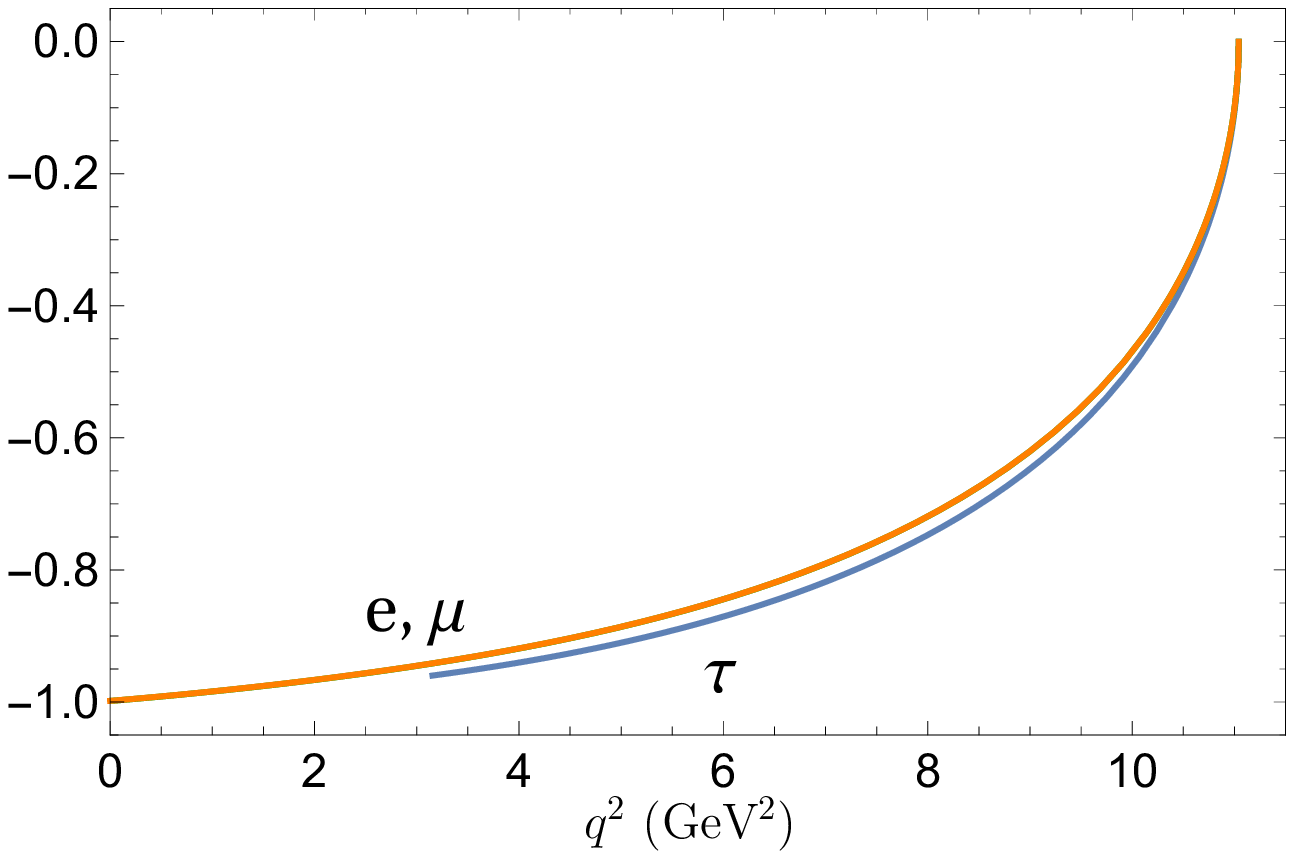}\ \
 \  \includegraphics[width=8cm]{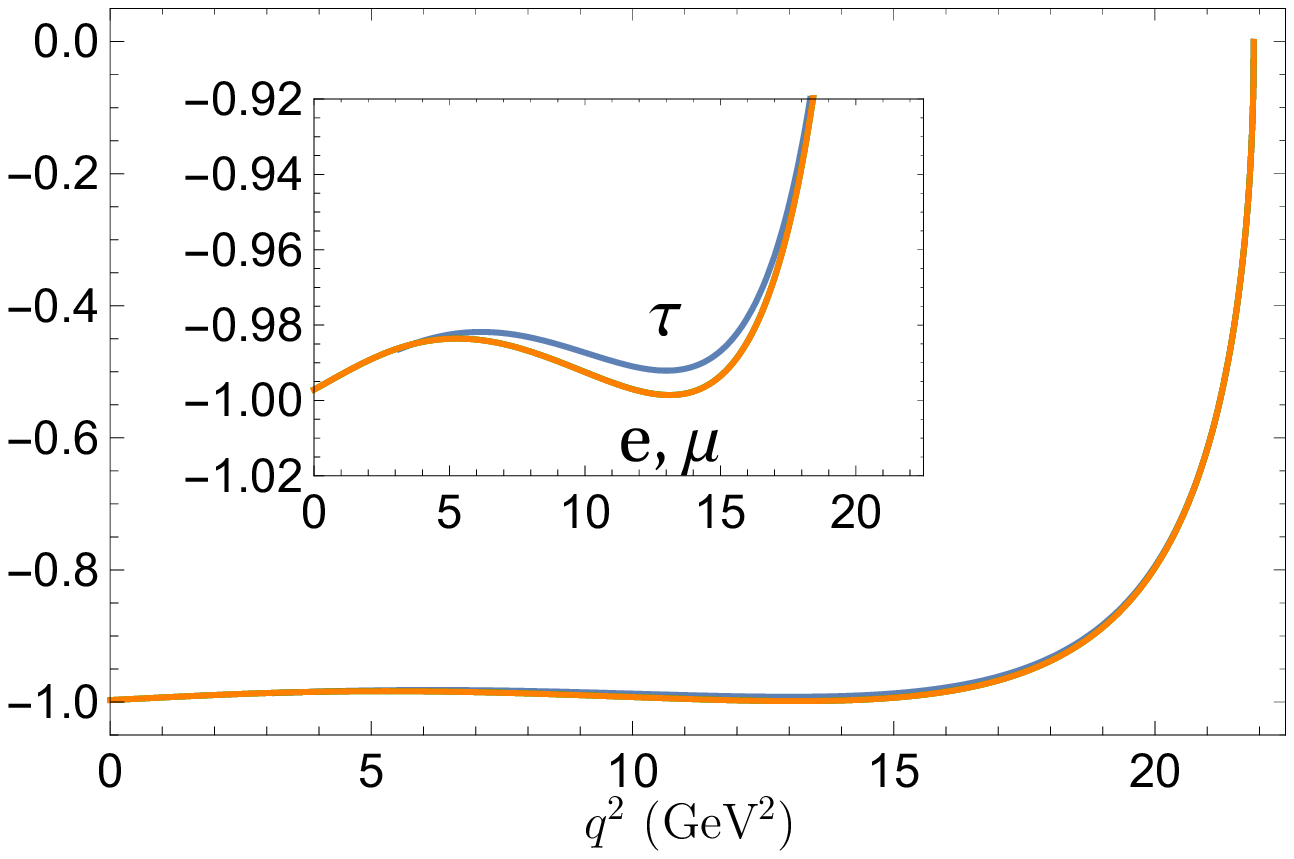}

  \caption{Predictions for the longitudinal polarization $P_L(q^2)$ of the final baryon  in the
    $\Xi_b\to \Xi_c\ell\nu_\ell$ (left) and $\Xi_b\to \Lambda\ell\nu_\ell$ (right)
    semileptonic decays. }
  \label{fig:PLXib}
\end{figure}

\begin{figure}
  \centering
 \includegraphics[width=8cm]{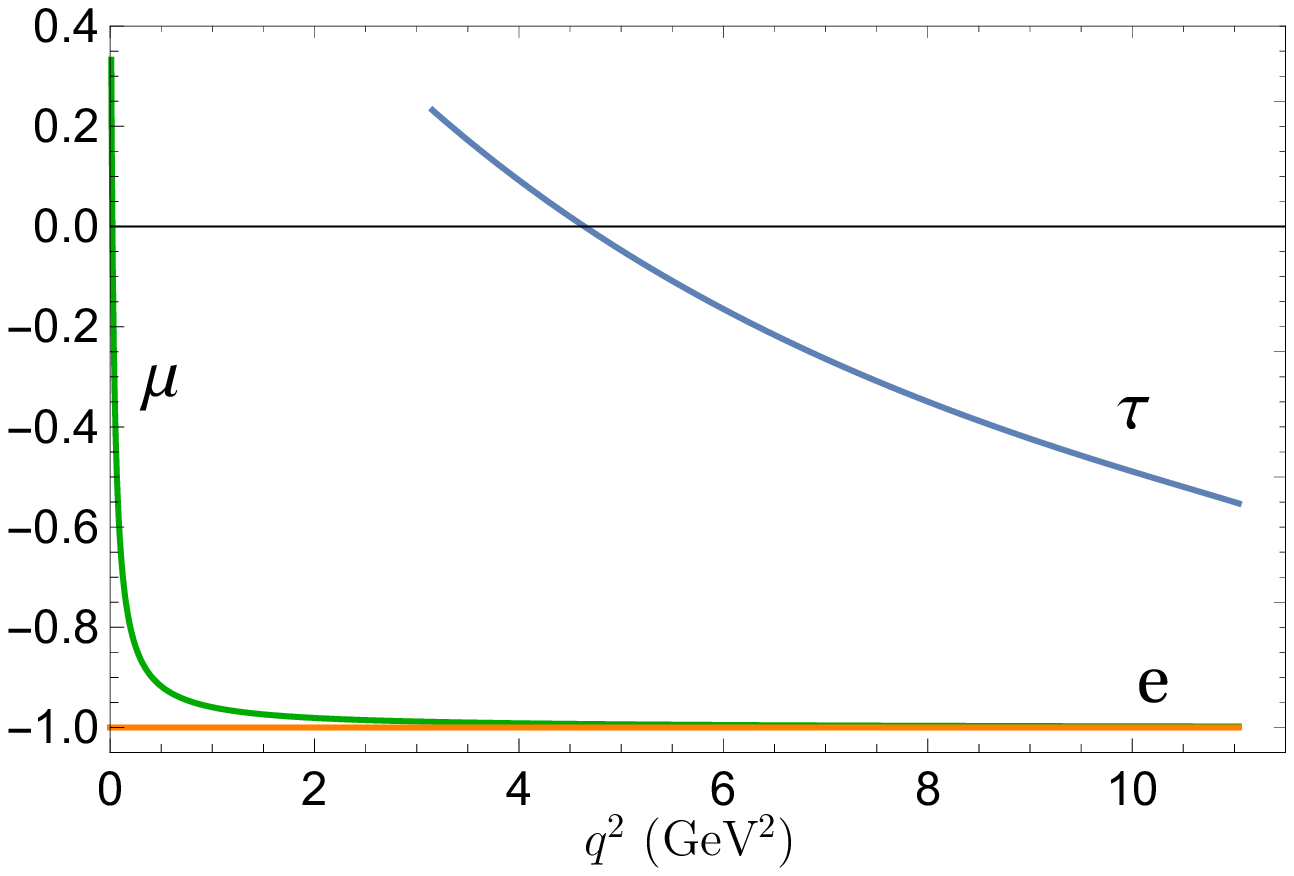}\ \
 \  \includegraphics[width=8cm]{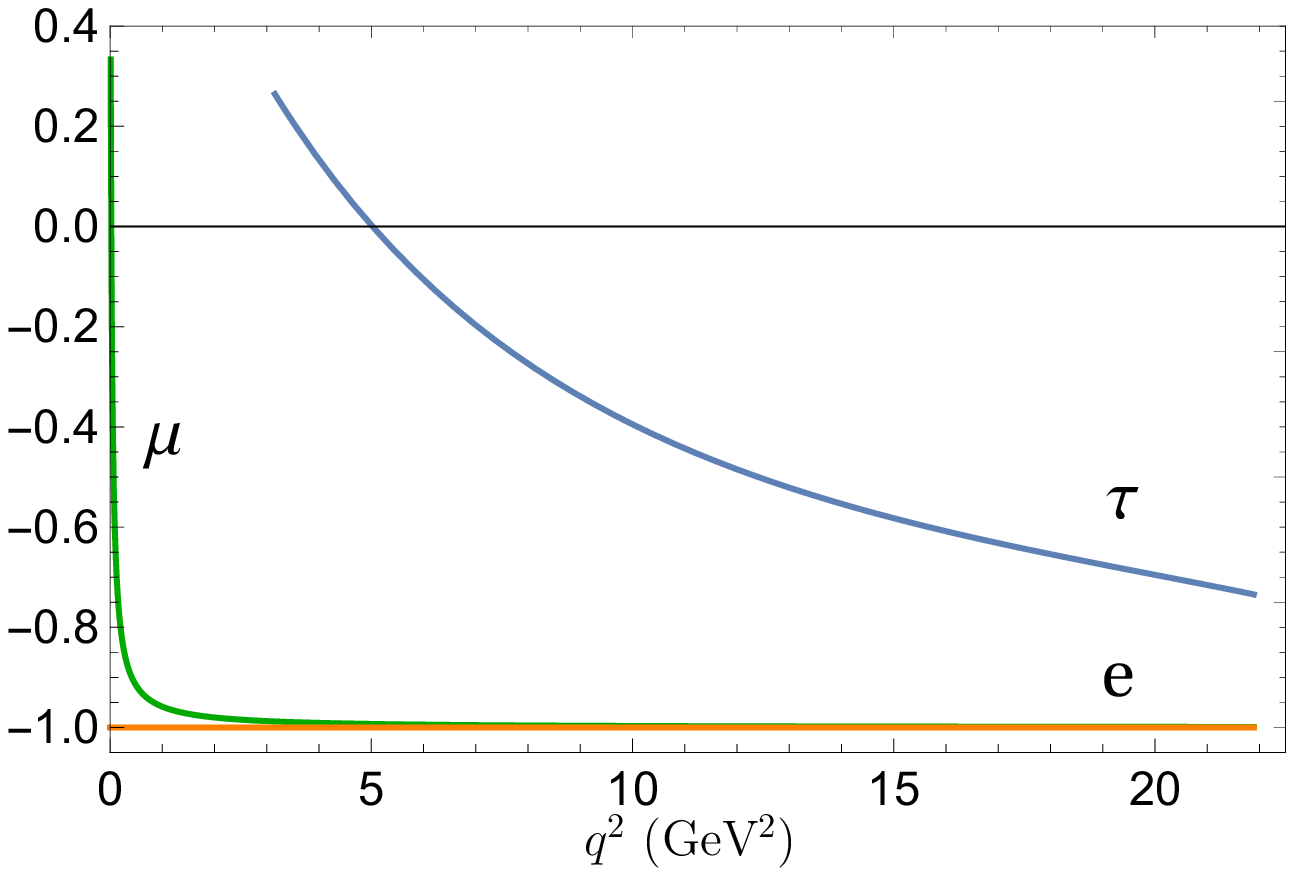}

  \caption{Predictions for the longitudinal polarization $P_\ell(q^2)$
    of the charged lepton  in the
    $\Xi_b\to \Xi_c\ell\nu_\ell$ (left) and $\Xi_b\to \Lambda\ell\nu_\ell$ (right)
    semileptonic decays. }
  \label{fig:PellXib}
\end{figure}

Many important observables are expressed in terms of the
helicity structures (\ref{eq:hhs}) (see \cite{giklsh} for details):\\
a) The forward-backward asymmetry of the charged lepton 
\begin{equation}
  \label{eq:afb}
  A_{FB}(q^2)=\frac{\frac{d\Gamma}{dq^2}({\rm forward})-\frac{d\Gamma}{dq^2}({\rm backward})}{\frac{d\Gamma}{dq^2}}
=-\frac34\frac{{\cal H}_P(q^2)+2\frac{m_\ell^2}{q^2}{\cal H}_{SL}(q^2)}{{\cal H}_{tot}(q^2)}.\qquad
\end{equation}
b) The convexity parameter  
\begin{equation}
  \label{eq:cf}
  C_F(q^2)=\frac34\left(1-\frac{m_\ell^2}{q^2}\right)\frac{{\cal H}_U(q^2)-2{\cal H}_L(q^2)}{{\cal H}_{tot}(q^2)}.
\end{equation}
c) The longitudinal polarization of the final baryon $\Xi_c(\Lambda)$ 
\begin{equation}
P_L(q^2)=\frac{[{\cal H}_P(q^2)+{\cal H}_{L_P}(q^2)]\left(1+\frac{m_\ell^2}{2q^2}\right)+3 \frac{m_\ell^2}{2q^2}{\cal H}_{S_P}(q^2)}{{\cal H}_{tot}(q^2)}.
\end{equation}
d) The longitudinal polarization of the charged lepton $\ell$ 
\begin{equation}
P_\ell(q^2)=-\frac{{\cal H}_U(q^2)+{\cal H}_L(q^2)-\frac{m_\ell^2}{2q^2}[{\cal H}_U(q^2)+{\cal H}_L(q^2)+3{\cal H}_{S}(q^2)]}{{\cal H}_{tot}(q^2)}.
\end{equation}
We plot these observables in Figs.~\ref{fig:afbXib}-\ref{fig:PellXib}
for heavy-to-heavy $\Xi_b\to\Xi_c\ell\nu_\ell$ and heavy-to-light
$\Xi_b\to \Lambda\ell\nu_\ell$ semileptonic decays. Our predictions
for the decay rates, branching fractions and asymmetry parameters are
given in Table~\ref{drbr}.  The decay rates are calculated  using the CKM values
$|V_{cb}|=(3.90\pm0.15)\times 10^{-2}$, 
$|V_{ub}|=(4.05\pm0.20)\times10^{-3}$ extracted from our previous
analysis of the heavy $B$ and $B_s$ meson decays \cite{slbdecay}. The average values of the
forward-backward asymmetry of the charged lepton $\langle
A_{FB}\rangle$, the convexity parameter $\langle C_F\rangle$ and the
longitudinal polarization of the final baryon $\langle P_L\rangle$ and
the charged lepton $\langle P_\ell\rangle$
are calculated by separately integrating the numerators and
denominators over $q^2$.

\begin{table}
\caption{$\Xi_b$ semileptonic decay rates, branching fractions and
  asymmetry parameters. }
\label{drbr}
\begin{ruledtabular}
\begin{tabular}{cccccccc}
Decay& $\Gamma$ (ns$^{-1}$) & $\Gamma/|V_{qb}|^2$ (ps$^{-1}$)& $Br$ (\%)
  & $\langle A_{FB}\rangle$ &$\langle C_F\rangle$& $\langle P_L\rangle$& $\langle P_\ell\rangle$\\
\hline
$\Xi_b\to\Xi_ce\nu_e$ &39.1&25.7 & 6.15 & $0.199$&$-0.540$&$-0.794$&$-1$\\
$\Xi_b\to\Xi_c\mu\nu_\mu$ &39.0&25.6 & 6.13  &$0.194$&$-0.525$&$-0.794$&$-0.985$\\
 $\Xi_b\to\Xi_c\tau\nu_\tau$&12.7  &8.4 & 2.00 & $-0.018$& $-0.087$&$-0.703$&$-0.324$\\
$\Xi_b\to \Lambda e\nu_e$ &0.164&10.0 & 0.026  & $0.384$&$-0.226$&$-0.919$&$-1$\\
$\Xi_b\to \Lambda\mu\nu_\mu$ &0.164&9.99 & 0.026  & $0.382$&$-0.223$&$-0.919$&$-0.996$\\$\Xi_b\to \Lambda\tau\nu_\tau$&0.117  &7.17 & 0.018 & $0.213$& $-0.073$&$-0.903$&$-0.579$\\
\end{tabular}
\end{ruledtabular}
\end{table}

Since the discrepancy between predictions of the Standard Model and
experimental data in heavy meson semileptonic decays is observed for
the ratio of branching ratios of decays  involving 
$\tau$ and a muon or electron \cite{pdg,bifani}, it is 
important to investigate similar decays of the heavy baryons. Using
our results we get the following predictions for the ratios of the
$\Xi_b$ baryon branching fractions $(l=e,\mu)$
\begin{eqnarray}
  \label{eq:ratio}
  R_{\Xi_c}&=&\frac{Br(\Xi_b\to\Xi_c\tau\nu_\tau)}{Br(\Xi_b\to\Xi_c
  l\nu_l)}=0.325\pm 0.010,\cr
R_{\Lambda}&=&\frac{Br(\Xi_b\to \Lambda\tau\nu_\tau)}{Br(\Xi_b\to \Lambda
               l\nu_l)}=0.717\pm0.021.
\end{eqnarray}
Note that part of the theoretical uncertainties cancels in these ratios
and we roughly estimate them to be about 3\%. The values of these
ratios are slightly larger than the corresponding values $R_{\Lambda_c}=0.313$
and $R_p=0.649$ calculated previously for the semileptonic $\Lambda_b$
decays \cite{Lambdabsl}. Any significant
deviations from these results, if
observed, can be interpreted as a signal of the new physics contributions.

For the semileptonic $\Lambda_b$ decays the LHCb Collaboration
measured the ratio of the heavy-to-heavy and heavy-to-light decays
\cite{lhcb}. If we consider the similar ratio
for the $\Xi_b$ semileptonic decays, we get the following results   
\begin{eqnarray}\label{rxl}
  R_{\Xi_c\Lambda}^l&=&\frac{Br(\Xi_b\to \Lambda l\nu_l)}{Br(\Xi_b\to \Xi_c
  l\nu_l)}=(0.389\pm 0.012)\frac{|V_{ub}|^2}{|V_{cb}|^2}=(4.2\pm0.5)\times10^{-3}
  , \qquad (l=e,\mu),\cr
 R_{\Xi_c\Lambda}^\tau&=&\frac{Br(\Xi_b\to \Lambda \tau\nu_\tau)}{Br(\Xi_b\to \Xi_c
  \tau\nu_\tau)}=(0.854\pm 0.025)\frac{|V_{ub}|^2}{|V_{cb}|^2}=(9.2\pm1.2)\times10^{-3}.
\end{eqnarray}
The measurement of such  ratios can provide an additional determination of
the ratio of the CKM matrix elements  $|V_{ub}|$ and $|V_{cb}|$. The
final value was obtained for our preferred CKM values and the error
bar includes combined theoretical uncertainties in the ratio of decay branching
fractions as well as the uncertainties in the ratio of the CKM values.

In Table~\ref{compth} we compare our results for the semileptonic
decay $\Xi_b\to\Xi_c\ell\nu_\ell$ with other calculations
\cite{ahn,efg,dutta}.~\footnote{The results and references of the
  earlier predictions can be found in Ref.~\cite{efg}.} The
previous investigations \cite{ahn,efg,dutta} employed heavy quark effective
theory to study this heavy-to-heavy baryon transition. Both the infinitely
heavy quark limit and the first order $1/m_Q$ corrections were
considered. In our present calculations we perform all calculations
without application of the nonrelativistic or heavy
quark expansions. We find our total decay rates and
branching ratios to be somewhat lower and slight deviations in other observables.  Note that the central value of our ratio $R_{\Xi_c}$ is
approximately 1.3 times larger than in Ref.~\cite{dutta}. This can
be attributed to the completely relativistic treatment of the decays
in our study.

\begin{table}
\caption{Comparison of theoretical predictions for the
  $\Xi_b\to\Xi_c\ell\nu_\ell$ semileptonic decay rates, branching fractions and
  asymmetry parameters. The superscript corresponds to the lepton type
  $e$ or $\tau$.}
\label{compth}
\begin{ruledtabular}
\begin{tabular}{ccccc}
Observable& this paper&\cite{ahn}&\cite{efg} & \cite{dutta}\\
\hline
$\Gamma^e/|V_{cb}|^2$ (ps$^{-1}$)& 25.7&$29.6(2.5)$ & 31& \\
$Br^e$ (\%)& 6.15& &7.4&9.22\\
$Br^\tau$ (\%)&2.00& & &2.35\\
$R_{\Xi_c}$&0.325& & &0.255\\
$\langle A_{FB}^e\rangle$&0.199& & &0.163\\
$\langle A_{FB}^\tau\rangle$&$-0.018$& & &$-0.042$\\
$\langle C_F^e\rangle$&$-0.540$& & &$-0.697$\\
$\langle C_F^\tau\rangle$&$-0.087$& & &$-0.103$\\
$\langle P_L^e\rangle$&$-0.794$&$-0.820(4)$&$-0.802$\\
  $\langle P_e\rangle$&$-1$& & &$-1$\\
  $\langle P_\tau\rangle$&$-0.324$& & &$-0.317$\\  
\end{tabular}
\end{ruledtabular}
\end{table}

\section{Conclusion}

The form factors of the heavy-to-heavy $\Xi_b\to\Xi_c$ and
heavy-to-light $\Xi_b\to\Lambda$ weak transitions were calculated in
the relativistic quark-diquark picture with the comprehensive account
of the relativistic effects. The relativistic baryon wave functions
were used for the evaluation of the corresponding form factors. The form factor momentum transfer $q^2$ dependence was
explicitly determined in the whole accessible kinematical range
without extrapolations or additional model assumptions. It was found
that the analytic approximation for these form factors (\ref{fitff})
accurately reproduces their $q^2$ behaviour with the parameters listed
in Tables~\ref{ffXibXic}, \ref{ffXibLambda}. The helicity formalism
was used for the calculation of the differential and total
semileptonic decay rates and other useful observables which are given
in Table~\ref{drbr}. The ratios $R_{\Xi_c}$  and $R_{\Lambda}$
(\ref{eq:ratio}) of the semileptonic $\Xi_b$  to $\Xi_c$ and
$\Lambda$ decay rates involving $\tau$ and electron are obtained. Their measurement can serve as an additional test of the lepton
flavour universality. The measurement of the ratio (\ref {rxl}) of heavy-to-light
$\Xi_b\to\Xi_c \ell\nu_\ell$ to heavy-to-heavy $\Xi_b\to\Xi_c
\ell\nu_\ell$ semileptonic decays can provide the independent determination of the CKM
ratio $|V_{ub}|/|V_{cb}|$. To our knowledge this is the first detailed
study of the CKM suppressed semileptonic $\Xi_b\to\Lambda\ell\nu_\ell$ decays.

\acknowledgments
The authors are grateful to D. Ebert and M. Ivanov  for valuable  discussions.


\begin{thebibliography}{00}
\bibitem{pdg} 
  M.~Tanabashi {\it et al.} [Particle Data Group],
  ``Review of Particle Physics,''
  Phys.\ Rev.\ D {\bf 98}, no. 3, 030001 (2018).
\bibitem{bifani} 
  S.~Bifani, S.~Descotes-Genon, A.~Romero Vidal and M.~H.~Schune,
  ``Review of Lepton Universality tests in $B$ decays,''
  arXiv:1809.06229 [hep-ex].
\bibitem{lhcbjp} 
  R.~Aaij {\it et al.} [LHCb Collaboration],
  ``Measurement of the ratio of branching fractions $\mathcal{B}(B_c^+\,\to\,J/\psi\tau^+\nu_\tau)$/$\mathcal{B}(B_c^+\,\to\,J/\psi\mu^+\nu_\mu)$,''
  Phys.\ Rev.\ Lett.\  {\bf 120}, no. 12, 121801 (2018).
 \bibitem{tiks} 
  C.~T.~Tran, M.~A.~Ivanov, J.~G.~Körner and P.~Santorelli,
  ``Implications of new physics in the decays $B_c \to (J/\psi,\eta_c)\tau\nu$,''
  Phys.\ Rev.\ D {\bf 97}, no. 5, 054014 (2018).
\bibitem{Lambdabsl} 
  R.~N.~Faustov and V.~O.~Galkin,
  ``Semileptonic decays of $\Lambda_b$ baryons in the relativistic quark model,''
  Phys.\ Rev.\ D {\bf 94}, no. 7, 073008 (2016).
 \bibitem{giklsh} 
T.~Gutsche, M.~A.~Ivanov, J.~G.~K\"orner, V.~E.~Lyubovitskij, P.~Santorelli and N.~Habyl,
  ``Semileptonic decay $\Lambda_b \to \Lambda_c + \tau^- + \bar{\nu_\tau}$ in the covariant confined quark model,''
  Phys.\ Rev.\ D {\bf 91}, no. 7, 074001 (2015)
  Erratum: [Phys.\ Rev.\ D {\bf 91}, no. 11, 119907 (2015)].
  \bibitem{hbarregge} 
  D.~Ebert, R.~N.~Faustov and V.~O.~Galkin,
  ``Spectroscopy and Regge trajectories of heavy baryons in the relativistic quark-diquark picture,''
  Phys.\ Rev.\ D {\bf 84}, 014025 (2011).
\bibitem{sbar}
  R.~N.~Faustov and V.~O.~Galkin,
  ``Strange baryon spectroscopy in the relativistic quark model,''
  Phys.\ Rev.\ D {\bf 92}, no. 5, 054005 (2015).
\bibitem{bcmass}
  D.~Ebert, V.~O.~Galkin and R.~N.~Faustov,
   ``Properties of heavy quarkonia and $B_c$ mesons in the relativistic quark model,''
   Phys. Rev. D {\bf 67}, 014027 (2003);
   ``Spectroscopy and Regge trajectories of heavy quarkonia and $B_c$ mesons,''
   Eur.\ Phys.\ J.\ C {\bf 71}, 1825 (2011).
 \bibitem{hmmass} 
  D.~Ebert, R.~N.~Faustov and V.~O.~Galkin,
  ``Heavy-light meson spectroscopy and Regge trajectories in the relativistic quark model,''
  Eur.\ Phys.\ J.\ C {\bf 66}, 197 (2010).
 \bibitem{slbdecay} 
  D.~Ebert, R.~N.~Faustov and V.~O.~Galkin,
  ``Analysis of semileptonic $B$ decays in the relativistic quark model,''
  Phys.\ Rev.\ D {\bf 75}, 074008 (2007);
``Rare $B\to \pi l\bar l$ and $B\to\rho l \bar l$ decays in the relativistic quark model,''
Eur.\ Phys.\ J.\ C {\bf 74}, no. 6, 2911 (2014).
\bibitem{lhcb} 
  R.~Aaij {\it et al.} [LHCb Collaboration],
  ``Determination of the quark coupling strength $|V_{ub}|$ using baryonic decays,''
  Nature Phys.\  {\bf 11}, 743 (2015).
  \bibitem{ahn} 
  C.~Albertus, E.~Hernandez and J.~Nieves,
  ``Nonrelativistic constituent quark model and HQET combined study of semileptonic decays of $\Lambda_b$ and $\Xi_b$ baryons,''
  Phys.\ Rev.\ D {\bf 71}, 014012 (2005).
   \bibitem{efg} 
  D.~Ebert, R.~N.~Faustov and V.~O.~Galkin,
  ``Semileptonic decays of heavy baryons in the relativistic quark model,''
  Phys.\ Rev.\ D {\bf 73}, 094002 (2006).
  \bibitem{dutta} 
  R.~Dutta,
  ``Phenomenology of $\Xi_b \to \Xi_c\,\tau\,\nu$ decays,''
  Phys.\ Rev.\ D {\bf 97}, no. 7, 073004 (2018).
 
  
\end{thebibliography}
\end{document}